\documentclass[letterpaper,twocolumn,10pt]{article}
\usepackage{usenix,epsfig,endnotes}
\usepackage{times}
\usepackage{enumitem}
\usepackage{graphicx}
\usepackage{tabularx}
\usepackage{amsmath}
\usepackage[sort]{natbib}
\usepackage{hhline}
\usepackage[font=small,labelfont=bf]{caption}
\usepackage{subcaption}
\usepackage{slashbox}
\usepackage{todonotes}
\usepackage[hyphens]{url}
\hypersetup{colorlinks=true,breaklinks=true}
\usepackage{algorithm}
\usepackage{algorithmicx}
\usepackage{algpseudocode}
\usepackage{blindtext}
\usepackage{diagbox}
\usepackage{booktabs}
\usepackage{xcolor}
\usepackage{subcaption}
\usepackage{hyphenat}
\usepackage[group-separator={,},group-minimum-digits=4]{siunitx}
\usepackage{authblk}

\setlist{nosep}

\usepackage[compact]{titlesec}
\titlespacing*{\paragraph}{0pt}{.5ex plus .2ex minus .2ex}{6pt}

\usepackage{amssymb}

\renewcommand{\P}{\mathbb{P}}

\newcommand{\Prob}[1]{\P\!\left[ {#1} \right]}

\usepackage{array}
\newcolumntype{L}[1]{>{\raggedright\let\newline\\\arraybackslash\hspace{0pt}}m{#1}}
\newcolumntype{C}[1]{>{\centering\let\newline\\\arraybackslash\hspace{0pt}}m{#1}}
\newcolumntype{R}[1]{>{\raggedleft\let\newline\\\arraybackslash\hspace{0pt}}m{#1}}

\hypersetup{
    colorlinks,
    citecolor=black,
    filecolor=black,
    linkcolor=black,
    urlcolor=black
}

\begin{document}

\newcommand{\company}{Facebook\xspace}
\newcommand{\ZippyDB}{ZippyDB\xspace}
\newcommand{\system}{DIRECT\xspace}
\newcommand{\Mod}[1]{\ (\text{mod}\ #1)}

\newcommand{\unsure}[2][1=]{\todo[linecolor=red,backgroundcolor=red!25,bordercolor=red,#1]{#2}}
\newcommand{\change}[2][1=]{\todo[linecolor=blue,backgroundcolor=blue!25,bordercolor=blue,#1]{#2}}

\date{}

\title{\Large \bf Live Recovery of Bit Corruptions in Datacenter Storage Systems}

\author[1,2]{Amy Tai}
\author[2]{Andrew Kryczka}
\author[2]{Shobhit Kanaujia}
\author[2]{Chris Petersen}
\author[2]{Mikhail Antonov}
\author[2]{Muhammad Waliji}
\author[1]{Kyle Jamieson}
\author[1]{Michael J. Freedman}
\author[3]{Asaf Cidon}
\affil[1]{Princeton University}
\affil[2]{Facebook, Inc.}
\affil[3]{Stanford University and Barracuda Networks}

\maketitle


\subsection*{Abstract}
Due to its high performance and decreasing cost per bit, 
flash is becoming the main storage medium in datacenters for hot data.
However, flash endurance is a perpetual problem, and due to technology
trends, subsequent generations of flash devices
exhibit progressively shorter lifetimes
before they experience uncorrectable bit errors.

In this paper we propose extending flash lifetime by allowing
devices to expose higher bit error rates.
To do so, we present \system, a novel set of policies that leverages latent
redundancy in distributed storage systems
to recover from bit corruption errors with minimal performance
and recovery overhead.
In doing so, \system can significantly extend the lifetime of flash
devices by effectively utilizing these devices even after they begin exposing
bit errors.

We implemented \system on two real-world storage systems:
\ZippyDB, a distributed key-value store backed by RocksDB, and
HDFS, a distributed file system.
When tested on production traces at \company, \system
reduces application-visible error rates in \ZippyDB by more than $10^2$
and recovery time by more than $10^4$.
\system also allows HDFS to tolerate a $10^4$--$10^5$
higher bit error rate without experiencing application-visible errors.

  \setlength{\belowcaptionskip}{-10pt}
\section{Introduction}

Flash is rapidly becoming the dominant storage medium for hot data in
datacenters~\cite{facebookflash,googleflash}, since it offers
significantly lower latency and higher throughput than
hard disks.  Many storage systems are built
atop flash, including databases~\cite{leveldb, voldemort, rocksdb-facebook,
flash-hbase}, caches~\cite{nitro,pannier,RIPQ,lightning}, and file
systems~\cite{HDFS-ssd,SDF}.

However, a perennial problem of flash is its limited endurance,
or how long it can reliably correct raw bit errors.
As device writes are the main contributor to flash wear,
this lifetime is measured in the number of writes or program-erase (P/E) cycles
the device can tolerate before exceeding an uncorrectable bit error 
threshold. Uncorrectable bit errors are device errors that are exposed to the application
and occur when there are too many raw bit errors for the device to correct.

\begin{figure*}[t]
  \begin{subfigure}[t]{0.48\linewidth}
  \centering
    \includegraphics[width=\textwidth]{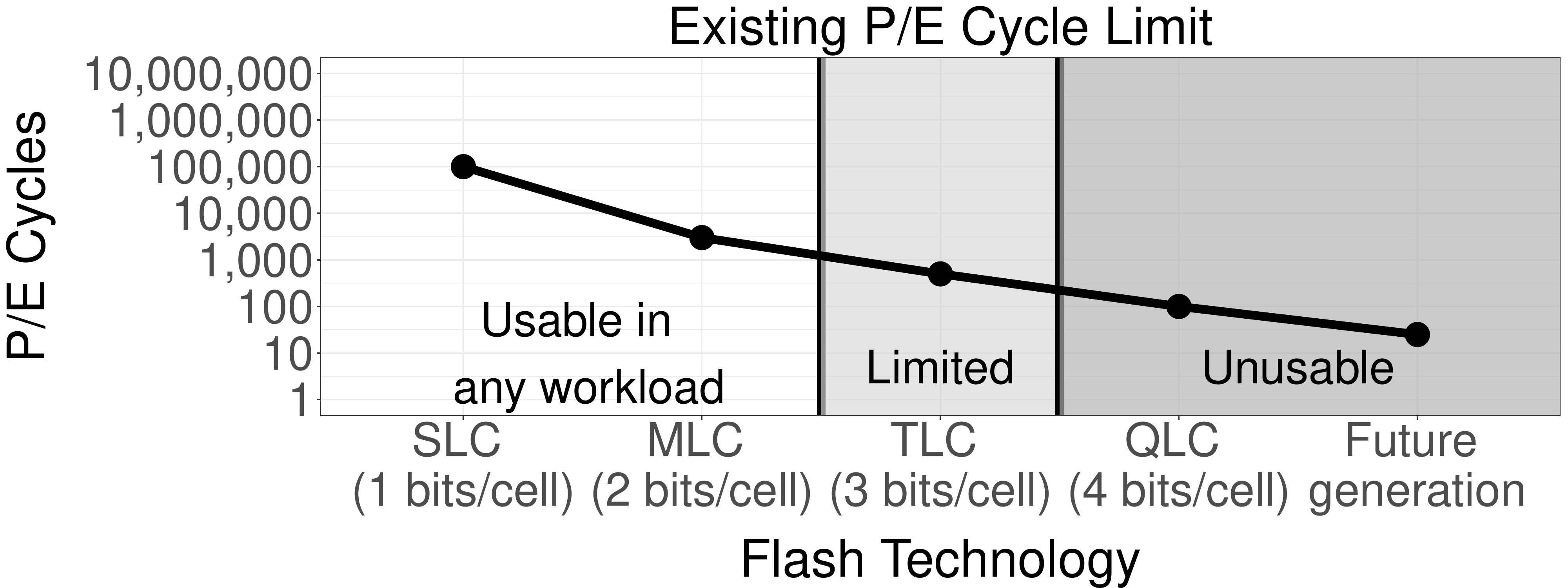}
    \caption{Existing hardware-based error correction.}
    \label{fig:before}
  \end{subfigure} \hfill
  \begin{subfigure}[t]{0.48\linewidth}
  \centering
    \includegraphics[width=\textwidth]{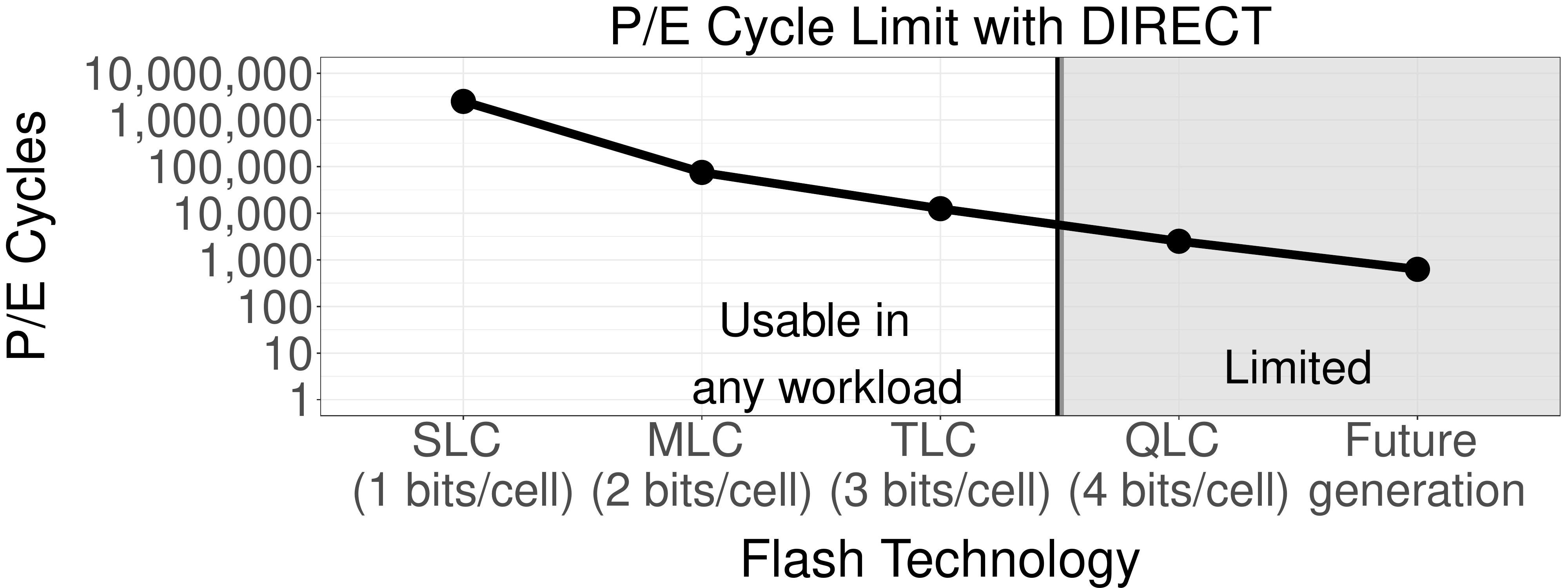}
    \caption{Augmenting existing error correction with \system.} 
    \label{fig:after}
  \end{subfigure}
\caption{For each generation of flash bit density, the average number of P/E cycles after
  which the uncorrectable bit error rate falls below the manufacturer specified level ($10^{-15}$).
  Beyond MLC, flash becomes constrained to read-heavy applications~\cite{caimutlu}.
 With current hardware-based error correction, with QLC technology and beyond, flash becomes almost unusable~\cite{tomshardware,qlc,flashmemorysummit}.
 \system enables the adoption of denser flash technologies because errors can be
 handled by the distributed storage application. The uncorrectable bit error rate that can be tolerated by \system was computed
 using the model from \S\ref{sec:availability-analysis}, while the uncorrectable bit error rate to 
 P/E conversion was computed using data from a Google study~\cite{googleflash}.} 
\end{figure*}

In hyper-scale datacenter environments, operators constantly seek to
reduce flash wear by 
limiting flash writes~\cite{albrecht2013janus,facebookflash}.
At \company for example,
a dedicated team monitors application flash writes to ensure they do not
prematurely exceed manufacturer-defined device lifetimes.
To make matters worse, each subsequent flash generation tolerates a smaller number of writes before 
reaching end-of-life (see Figure~\ref{fig:before})~\cite{bleak}.
Further, given the scaling challenges of DRAM~\cite{lee2016technology,kang2014co},
and the increasing cost gap between DRAM and flash~\cite{dram-climb,dramPrices}, 
many operators are migrating services from DRAM to flash~\cite{mcdipper,NVM-DRAM}.

There is a variety of work that attempts to extend flash lifetime
by delaying the onset of bit errors~\cite{Tanakamaru20112,cai2015data,retentionrelaxation,LDPC-SSD,wom-codes,jeong2014lifetime,
wisckey,SILT,rocksdb,leveldb}.
This paper takes the opposite approach.
We observe that flash endurance can be extended
by \emph{allowing} devices to go beyond
their advertised uncorrectable bit error rate (UBER) and
embracing the use of flash disks at much higher error
rates. To do so however, distributed storage systems must be
retrofitted with a new paradigm that does \emph{not}
assume corruption-free devices. Google recently
released a whitepaper suggesting a similar approach~\cite{brewerdisks}.

Traditionally, distributed storage systems are built to tolerate machine or disk failures, not
bit corruption on an individual data block. To recover from machine failures,
storage systems re-replicate an entire server,
but such heavy-handed recovery is inappropriate for handling
errors that may affect only a single bit.
Instead, our key insight is
that minimizing \emph{error amplification},
or the number of bits needed to recover a bit error,
enables us to use corruption-prone devices by
reducing the probability of application-visible errors
and improving recovery performance.

We introduce Distributed error Isolation and RECovery Techniques (DIRECT), which is rooted
in the observation that (1) datacenter storage systems replicate data on remote
servers, and (2) this redundancy can correct bit error rates orders of magnitude beyond
the hardware error correction mechanisms implemented on the device.
\system is a set of three simple general-purpose policies that,
when implemented, enable distributed storage systems to
achieve high availability and
correctness in the face of uncorrectable bit errors:
\begin{enumerate}
\item \textbf{Minimize error amplification.}
\system detects errors using existing error detection mechanisms (e.g., checksums)
and recovers data from remote servers at the smallest possible granularity.
\item \textbf{Local metadata protection.}
To recover from a corruption in local metadata (e.g., database index), often a large
amount of data must be re-replicated. \system avoids this by adding local redundancy to 
local metadata. 
\item \textbf{Safe recovery semantics.}
Any recovery operations on corrupted data must be serialized against
concurrent read and write operations with respect to the system's consistency guarantees.
\end{enumerate}

We design and implement the \system policies in two popular systems that
are illustrative of widely-used storage architectures:
(1) \ZippyDB,
a distributed key-value store used in production at \company and
backed by RocksDB, a popular storage engine based
on the log-structured merge tree~\cite{LSM}, and (2) the Hadoop
Distributed File System (HDFS), which is
representative of distributed storage systems that perform full-block replication.
In both systems, we minimize error amplification by isolating bit errors to data regions with sizes on the order of
kilobytes, making recovery very fast compared
to re-replication of an entire server.

\system enables HDFS to tolerate much higher bit error rates because
blocks in HDFS are immutable after write,
so \system fixes bit errors by comparing
across replicas of the same block (\S\ref{sec:sleddyhdfs}).
On the other hand, recovery is challenging in
RocksDB due to background compaction operations
and key-versioning. Compaction makes it difficult not only to find the corrupted region
on one replica in another replica (different servers store the same key-value pairs
in different files), but also
to ensure that the recovered key-value pairs have consistent versions.
\system must make use of the distributed layer in \ZippyDB
to solve both these problems (\S\ref{sec:zipRecovery}).

Applying \system results in 
significant end-to-end improvements:
it reduces application-visible error rates in \ZippyDB by more than 100$\times$, reduces
recovery time by 10,000$\times$,
and reduces CPU consumption by 20\%-49\%.
It enables HDFS to tolerate bit error rates that are 10,000$\times$-100,000$\times$ greater.

With these performance improvements,
\system can lead to significant increases in device lifetime,
because it maintains the same probability of application-visible errors
at much higher device UBERs (for the computation, see \S\ref{sec:availability-analysis}).
An estimate of lifetime increase is shown in Figure~\ref{fig:after};
we estimate the number of P/E cycles gained by running to higher UBERs from
a Google study~\cite{googleflash}.
Depending on the system parameters, \system can increase the lifetime of devices by 10-100$\times$.
This allows datacenter operators to replace flash devices less often and
adopt lower cost-per-bit flash technologies
that have lower endurance.
\system also provides the opportunity to rethink the design
of existing flash-based storage systems, which are brittle in the face of corruption errors.
Furthermore, while this paper focuses on flash,
\system's principles also apply in other storage mediums, including NVM and hard disks.

In summary, this paper makes several contributions:
\begin{enumerate}
\item We observe that flash lifetime can be extended by allowing 
devices to expose higher bit error rates.
\item We propose \system, general-purpose software policies that enable storage systems
to maintain performance and high availability in the face of high hardware bit error rates.
\item We design and implement \system in two representative storage systems, \ZippyDB and HDFS.
\item We demonstrate that \system significantly speeds up recovery time due to disk corruptions,
and significantly lowers application-observable
errors in the resulting systems, allowing them to tolerate much higher hardware bit error rates.
\end{enumerate}

\section{Motivation}

\paragraph{What Limits Flash Endurance?}
Flash chips are composed of memory cells,
each of which stores an analog voltage value. The flash
controller reads the value stored in a certain memory cell by sensing
the voltage level of the cell and applying
quantization to determine the discrete value in bits.
The more bits stored in a cell, the narrower the voltage
range that maps to each discrete bit, so more precise voltage
sensing is
required to get a correct read.
Unfortunately, 
one of the primary ways to reduce cost per bit is
to increase the number of bits per cell,
which means that even small 
voltage perturbations can result in a misread.

Multiple factors cause voltage drift in a flash cell.
The dominant source, especially
in datacenter settings where most data is ``hot,'' is the
program\hyp{}erase (P/E) cycle, which involves applying a large high
voltage to the cell in order to drain its stored charge, thus wearing
the insulating layer in the flash cell~\cite{cai2015data}.  This increases the voltage drift in
subsequent values in the cell, which gradually leads to bit errors.  

3D NAND is a recent technology that has been adopted for further increasing
flash density by stacking cells vertically.
While 3D NAND relaxes physical limitations of 2D NAND (traditional flash) by enabling vertical stacking,
3D NAND inherits the reliability problems of 2D NAND, and further exacerbates them,
since a cell in 3D NAND has more adjacent (vertical) neighbors. For example, voltage retention
is worse, because voltage can now leak in three dimensions~\cite{3dnand-1,3dnand-book}.
Similarly, disturb errors that occur when 
adjacent cells are read or programmed
are also exacerbated~\cite{3dnand-reliability,kim2017evolution}.

\paragraph{Existing Hardware Reliability Mechanisms.}

To correct bit errors,
flash devices use error correcting codes (ECC), which are
implemented in hardware.
After the ECC pass, there could still be incorrect bits on the page. To address
these errors, SSDs also employ internal RAID across the dies
inside the flash device~\cite{micronRAIN, RAISE}.
After applying coding and RAID within the device, there will remain
a certain rate of \emph{uncorrectable bit errors} (UBER).
Together, ECC and internal RAID mechanisms can drive the error rates of SSDs from the raw
bit error rate of around $10^{-6}$ down to the $10^{-17}$ to $10^{-20}$ UBER range typical of
enterprise SSDs~\cite{sandisk}. ``Commodity'' SSD devices typically guarantee an UBER of $10^{-15}$.

However, the level of RAID striping is constant across generations,
because the number of dies inside a flash device remains constant. This
means that the corrective power of RAID is fixed. While it is possible
to create stronger ECC engines, the higher the
corrective power of the ECC, the more costly the device due to the 
complexity of the ECC circuit~\cite{flashMemPres,qlcECCEngine}.

\paragraph{Implications of Limited Flash Endurance.}

Flash technology has already reached the point where its
endurance is inhibiting its adoption and operation in various
datacenter use cases.
For example, QLC was recently introduced as the next generation flash cell technology.  However, it can
only tolerate 100-200 P/E
cycles~\cite{tomshardware,qlc,flashmemorysummit}, so it can only be
used for read-heavy use cases.  Datacenter applications that deal with
hot data, such as databases and analytics, typically need to update
objects frequently.  This has limited the adoption of QLC (and is
the reason that \company has avoided QLC flash).  Subsequent
cell technology generations will suffer from even greater problems.
Second, operational issues often dictate a device's usage lifetime.
While flash manufacturers are conservative with their
flash device lifetimes~\cite{googleflash}, 
flash is still only used for its advertised lifetime 
to simplify operational complexity. Further, in a hyper-scale datacenter
where it is common to source devices from multiple vendors,
the most conservative estimate of device lifetime across vendors is typically chosen
as the lifetime for a fleet of flash devices,
so that the entire fleet can be installed
and removed together.
However, if the distributed storage layer could tolerate much higher device error rates,
then datacenter operators would no longer have to make conservative
and wasteful estimates about entire fleets of flash devices. 

Third, because of the increase in DRAM prices due to its scaling challenges and tight supply~\cite{dramPrices,lee2016technology,kang2014co,dram-climb},
datacenter operators are migrating services from DRAM to flash~\cite{mcdipper,NVM-DRAM}.
This means that flash will be responsible for
many more workloads, further exacerbating the flash endurance problem.
Limited flash lifetime is already a problem in the datacenter, where operators
must limit applications to a certain write throughput per day to 
prevent prematurely wearing out a device. 

\begin{figure*}[t]
\begin{subfigure}{\columnwidth}
\includegraphics[trim={1.3cm 5.6cm 2.5cm 5.8cm},clip,scale=0.53,page=7]{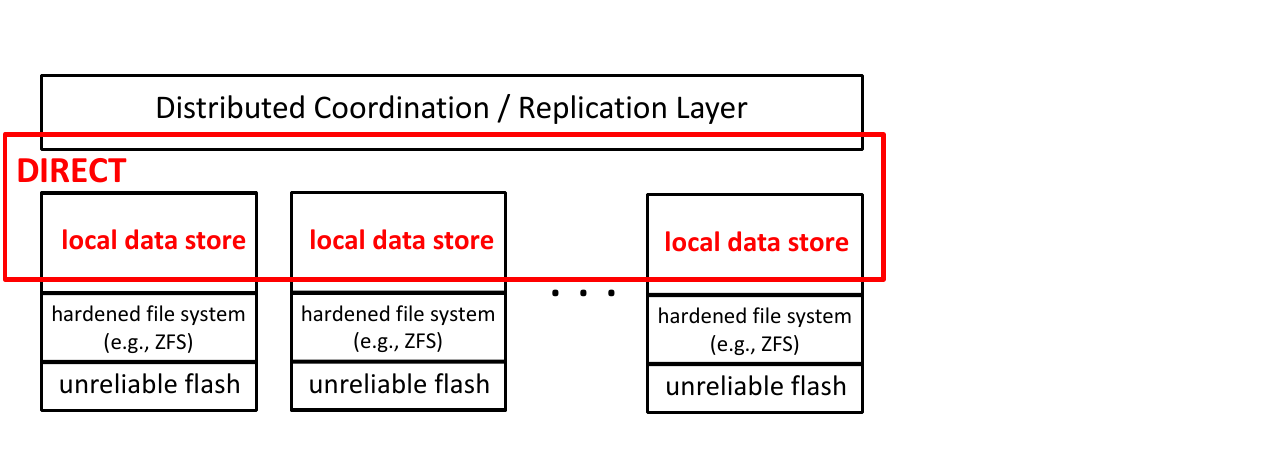}
\caption{}
\label{fig:SLED}
\end{subfigure}
\begin{subfigure}{\columnwidth}
\includegraphics[trim={1.3cm 2.8cm 3cm 8.8cm},clip,scale=0.52,page=15]{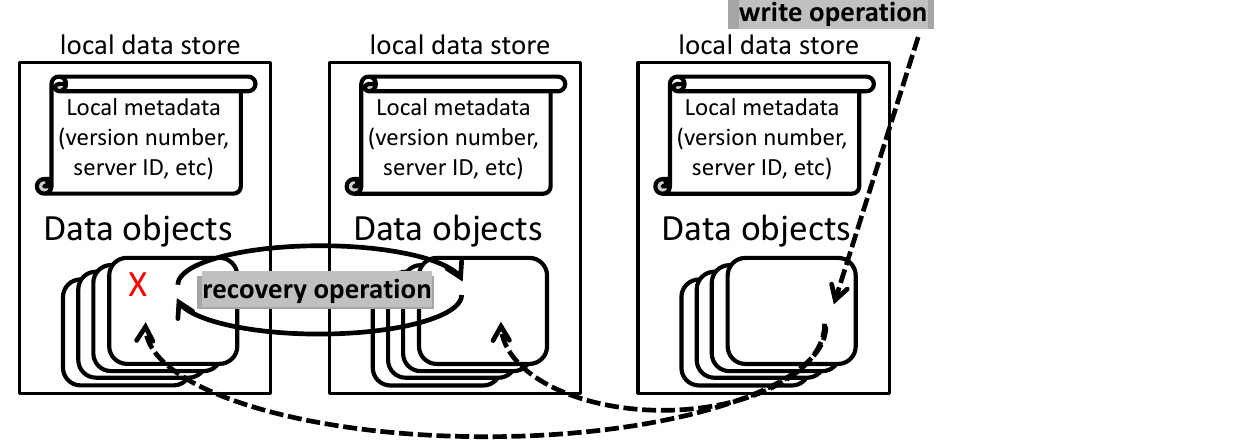}
\caption{}
\label{fig:localStore}
\end{subfigure}
\caption{(a) \system 
instruments cooperation between the local data stores
and the distributed coordination layer to fix errors in the local data store.
(b) Within the local data store,
bit errors can affect either data objects or metadata.
There must be precise semantics that define how recovery operations fixing data objects
interact with write operations.}
\label{fig:everything}
\end{figure*}

\section{\system Design}
\label{sec:design}

\system is a set of policies that
enables a
distributed storage system to maintain high availability
and correctness in the face of a high UBER.
We define
a distributed storage system
as a set of many local stores coupled with a distributed protocol layer that
replicates data and coordinates between the local stores.
Figure~\ref{fig:SLED} shows the \system storage stack, which accommodates unreliable flash
(flash that exposes high UBERs).
There is existing work on how to make local file systems tolerate
corruption errors (we survey some of these systems in \S\ref{sec:file-systems}).
However, there is no existing work on how to enable distributed storage systems,
or even local key-value stores, to tolerate bit corruption in a live production environment.
\system addresses these challenges.

\subsection{High Availability}
\label{sec:availability-analysis}

Within the local data store, bit errors affect either application data or application metadata,
as shown in Figure~\ref{fig:localStore}.
Maintaining multiple copies of each piece of data is the easiest
way for a system to recover from bit errors. 
Our observation is that 
this redundancy already exists for application data!

\paragraph{Distributed Redundancy.}
Distributed storage systems typically 
use replication~\cite{hadoop-replication} or erasure coding~\cite{HDFS-RAID,huang2012erasure} to store
redundant copies of data.
Hot data, which is stored on flash storage, is typically
replicated to avoid the higher bandwidth
and CPU consumption associated with reconstructing erasure coded blocks~\cite{huang2012erasure}.
In addition, erasure coding is not used for storage applications 
requiring fine-grained data access such as RocksDB.
Since distributed storage systems assume storage devices correct device-level errors,
they do not currently use replicas to correct bit errors~\cite{redundancy},
even though this redundancy can significantly
boost bit error resilience.

Consider the following example. Suppose a data block
is replicated in each of the three data stores shown in Figure~\ref{fig:localStore}.
If the block has size $B$, and the uncorrectable bit error rate (UBER) is $E$,
then the expected number
of errors in the block will be $B \cdot E$.  Since the block is
replicated across $R$ different servers, the storage
application can recover the block from a remote server when an error
occurs in at most $R-1$ of its replicas.  In this case, the only way
that the storage system would encounter an application-observable read
error is when at least one error exists in \emph{each} of the copies of
the block. Therefore, the probability of an application-level read error can be expressed as:
\noindent
\begin{equation*}
	\Prob{error}= (1 - (1 - E)^{B})^{R}\approx (E \cdot B)^{R}
\end{equation*}
\noindent
where we assume $E \cdot B << 1$ and use a Taylor series approximation.

Then for an UBER of $E=10^{-15}$, a block size of $B=128~MB$ (typical
of distributed file systems), and a replication factor of $R=3$,
the probability of error is $10^{-18}$ (files are measured in bytes, while UBER is in
bits).  This effectively is three
orders of magnitude lower than the UBER of each local disk.

\begin{table}[t!]
  \centering
  \small
  \begin{tabular}{p{1.4cm}p{2.4cm}p{2.4cm}}
    \toprule
    & \multicolumn{2}{c}{\textit{Probability of Application-Observable Error}} \\
    \textbf{UBER} & \textbf{Block Recovery} & \textbf{Chunk Recovery} \\
    \midrule
    $10^{-10}$ & $1 \cdot 10^{-3}$ & $3\cdot 10^{-10}$ \\
    $10^{-15}$ & $1 \cdot 10^{-18}$ & $1\cdot 10^{-28}$ \\
    \bottomrule
  \end{tabular}
  \caption{Probability of application-observable error comparing block-by-block recovery to chunk-by-chunk recovery,
  with an UBER of $10^{-10}$, and $10^{-15}$. Finer granularity recovery provides significantly higher protection
  against corruptions.}
  \label{tab:eval_comparison}
\end{table}

However, with relatively large blocks, the probability of encountering at least one
error in all block replicas quickly increases as UBER increases. For example, for an UBER
of $E=10^{-10}$, the expected number of errors in a single block will
be $B \cdot E=0.1$.  Thus, the probability of error in this case will be
$\Prob{error} \approx 0.001$. 
We make the observation that reducing $E\cdot B$,
by reducing $B$, will dramatically
reduce the probability of error.

\paragraph{Minimizing Error Amplification.}
\system captures this intuition with
error amplification ($B$ in the previous example),
or the number of bytes required to recover a bit error.
\system observes that 
\emph{the lower the error amplification, the lower the probability of error 
and the faster recovery can occur}.  This similarly implies a shorter
period of time spent in degraded durability and thus higher
availability.

In the example above, suppose the system can recover data at a finer 
granularity, for example, at chunk size $C=64$~KB.
Then a read error would occur if all three replicas of the same \emph{chunk}
have at least one bit error.
The revised probability of read error is:
\noindent
\begin{equation*}
	\Prob{error}= 1 - (1 - (1 - (1 - E)^{C})^{R})^{\frac{B}{C}}
\end{equation*}
\noindent
Assuming $E \cdot C << 1$, 
Taylor series approximation leads to
$(1 - (1 - E)^{C})^{R}) \approx (E \cdot C)^{R}$,
and assuming this value is much smaller than $\frac{B}{C}$,
the probability of an application-observable error when
correcting chunk-by-chunk is:
\noindent
\begin{equation*}
	\Prob{error} \approx (E \cdot C)^{R} \cdot \frac{B}{C}
\end{equation*}
\noindent
When $C=64$~KB and $E=10^{-10}$,
this probability is $3 \cdot 10^{-10}$, which
is much lower than the probability
when recovering at the block level (see Table~\ref{tab:eval_comparison}).

In HDFS, chunk recovery is precisely what allows \system to
tolerate higher bit error rates. The RocksDB data format is 
more complicated than the block format discussed in this section,
but \system also
isolates errors to data blocks ($\sim8$ KB) in RocksDB,
and this is responsible for significant improvements in recovery time.

\paragraph{Metadata Error Amplification.}

So far, we have discussed the effect of errors on data blocks. However,
error amplification can be even more severe if the error
occurs in local metadata.
For example, a corrupted local key-value store index
can prevent a data store from starting up, which can mean re-replication of
hundreds of GBs of data.
Even though the likelihood of errors in metadata is statistically lower than in data blocks
(metadata typically takes up much less space than data),
it requires stronger local protection to minimize
error amplification.
To address this problem, \system either locally duplicates
metadata or applies local software error correction.

\subsection{Correctness}
Minimizing error amplification of data blocks and correcting
data from remote replicas enables performant, live recovery of corrupted data blocks.
However, \system must also ensure
recovery operations preserve the correctness of the distributed storage system, which might
be dealing with concurrent write and read operations.
For example, in Figure~\ref{fig:localStore}, after both recovery operation and write operation,
 the corrupted data block
is both fixed and has the ``correct'' data with respect to consistency guarantees of
the system. 

Correct recovery is particularly difficult in RocksDB because of
key versioning.
The versions of the corrupted key-value pairs \emph{are not known},
because the corruption prevents the data from being read.
Hence in order to correctly recover corrupted key-value pairs, the system
must locate some consistent (up-to-date) version of each pair.
To do this, \system forces recovery operations to go through
the Paxos log in \ZippyDB, which can provide correct ordering (\S\ref{rocksProtection}).

\subsection{\system Policies}
To summarize, \system includes the following policies.

\begin{enumerate}
\item Systems must reduce error amplification
  of data objects and fix corruptions from remote replicas.
\item Systems must perform local metadata
  duplication to avoid high recovery costs from metadata corruption.
\item Systems must ensure safe recovery
  semantics.
\end{enumerate}

Note that the first and second policies apply exclusively to the local
data store and affect \emph{performance}, while
the third policy requires that the local data store interact with the
distributed coordination layer to ensure \emph{correctness} during recovery.

  \setlength{\belowcaptionskip}{-5pt}
\section{Implementing \system}
\label{sec:implementation}
To demonstrate the use of the \system approach, 
we integrate it into two systems:  \ZippyDB, a distributed key-value store
backed by RocksDB,
and HDFS, a popular distributed file system. 

\subsection{\ZippyDB-\system}
\subsubsection{\ZippyDB Overview}
\ZippyDB is a distributed key-value store used within \company that is backed by RocksDB
(i.e., RocksDB is the local data store in Figure~\ref{fig:SLED}).
\ZippyDB runs on tens of thousands of flash servers at \company, which makes it an ideal 
target for \system.
\ZippyDB provides a replication layer on top of RocksDB. \ZippyDB is logically separated into
shards, and each shard is fully replicated at least three ways.  Each
shard has a primary replica as well as a number of secondary replicas,
wherein each replica is backed by a separate RocksDB instance residing
on separate servers. Each
\ZippyDB server contains 100s of shards, including both primary and secondary replicas.
Hence, each \ZippyDB server actually contains
a large number of separate RocksDB instances.

\ZippyDB runs a Paxos-based protocol for shard operations to ensure consistency. The
primary shard acts as the leader for the Paxos entry, and each shard also has a Paxos log to 
persist each Paxos entry. Writes are considered durable when they are committed by
a quorum of shards, and write operations are applied to the local RocksDB store in the order that they are
committed. A separate service is responsible for monitoring the primary and triggering Paxos role changes.

\ZippyDB supports a variety of read consistencies depending on the client service: (1) strongly consistent reads,
which go through the primary; (2) read-after-write consistency, which can be served by any replica
if the client passes a Paxos entry to read-after; and (3) eventually consistent reads, which can go
to any replica.


\subsubsection{RocksDB Overview}
\begin{figure}[t]
\centering
\includegraphics[trim={6.6cm 5.5cm 13cm 4cm},clip,scale=0.5,page=9]{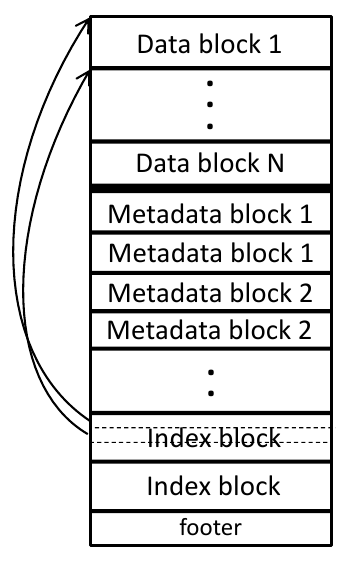}
\caption{RocksDB SST file format.
Index block entries point to keys in between data blocks, which means consecutive
index block entries will form a key range that contains all keys in the sandwiched data block.
\system writes each metadata block at least twice in-line (or uses an error correction code).}
\label{fig:newrockssst}
\end{figure}

RocksDB is a local key-value store that is based on a log-structured merge (LSM) tree~\cite{LSM}.
RocksDB batches writes in-memory---each write receives a sequence number
that enables key versioning---and flushes them into immutable files 
of sorted key-value pairs called sorted string table (SST) files. 
RocksDB SST files are composed of individually checksummed blocks, each of which can
be a data block or a metadata block. The metadata
blocks include index blocks that point to the keys at the start of each data block
(Figure~\ref{fig:newrockssst})~\cite{rocksdbSSTformat}.

SST files are organized into levels.
A key feature of RocksDB and other LSM tree-backed stores is background compaction, which periodically
scans SST files and compacts them into lower levels, as well as performs garbage collection on deleted
and overwritten keys.

\subsubsection{Implementing \system}
\label{rocksProtection}
 In \ZippyDB, 
if a compaction encounters a corruption, an entire server, which typically
has 100s of gigabytes to terabytes of data, will shutdown and attempt to drain its RocksDB shards to another machine.
Meanwhile, this sudden crash causes spikes in error rates and increases the load on other replicas
while the server is recovering.
To make matters worse, the new server could reside in a separate region,
further delaying time to recovery. All this leads to high error amplification:
a single bit error can cause the migration of terabytes of data.

\paragraph{Reducing Error Amplification of Data Blocks.}
We observe that
checksums in RocksDB are applied at the data block level, so
a data block is the smallest granularity at which a bit error can be recovered.
Data blocks are lists of key-value pairs, and key-value pairs are replicated at the \ZippyDB layer.
So if the metadata on an SST file
is correct (see below on how we protect per-SST file metadata),
a corrupted data block can be recovered
by fetching the pairs in the data block from another replica.
However, this is challenging for two reasons.

First, compactions are non-deterministic in RocksDB and
depend on a variety of factors such as available disk space and how compaction threads are scheduled. 
Hence, \emph{two replicas of the same RocksDB instance will have a different set of SST files}, making it
impossible to find an exact replica of the corrupted SST file, much less the corrupted data block.
Second, because the block is corrupted, it is impossible to know the exact key-value pairs that
were stored in that block.
Therefore, not only do we not know what data to look for on the other replica, we also don't know where to find it.

Instead of repairing the exact keys that are lost, we repair the corrupted data block by re-writing a larger
key range that covers the keys in the corrupted block.
The key range is determined from index blocks, which are a type of metadata block that
exist at the end of every SST file and record a key in the range between
consecutive data blocks, as shown in Figure~\ref{fig:newrockssst}. Hence, consecutive index block entries
form a key range which is guaranteed to contain the lost keys.

Unfortunately, just knowing the key range is not enough:
the existence of key versions in RocksDB and quorum replication in \ZippyDB compounds the problem.
In particular, a key must be recovered to a version greater than or equal to the lost key version, which could mean deleting it as key versions in RocksDB can be deletion markers.
Additionally, if we na\"{\i}vely fetch key versions from another replica, we may violate consistency.

\paragraph{Safe Recovery Semantics.}
\label{hihi}
To guide our recovery design, we introduce the following correctness requirement.
Suppose we learn from the index blocks that we must re-replicate key range $[a,b]$. This key range is requested from another replica, which assembles a set of fresh key-value pairs in $[a,b]$, which we call a patch.

\textbf{Safety Requirement}:
\emph{Immediately after patch insertion, the database must be in a state that reflects some prefix of the Paxos log.
Furthermore, this prefix must include the Paxos entries that originally updated the corrupted data block.}

In other words, patch insertion must bring \ZippyDB to some consistent state \emph{after} the versions of the corrupted keys; otherwise, if the patch inserts prior versions of the keys, then the database will appear to go backwards.

\begin{figure}[t]
\centering
\includegraphics[trim={1.4cm 9.5cm 14.5cm 6.5cm},clip,scale=0.8,page=10]{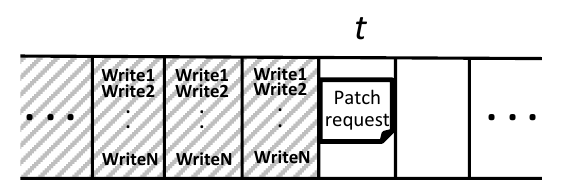}
\caption{To serialize a patch properly, we add it as a request in the Paxos log. If the patch request is serialized
at point $t$, then it must reflect all entries $t' < t$ (shaded). Furthermore, the patch request is not
batched with any writes to ensure atomicity.}
\label{fig:patchReq}
\end{figure}

Because the Paxos log serializes updates to \ZippyDB, the cleanest way to find a prefix to recover up to
is to serialize the patch insertion via the Paxos log. Then if patch insertion gets serialized as entry $t$ in the log,
the log prefix of the patch must reflect all Paxos entries $t' < t$, as shown in Figure~\ref{fig:patchReq}.
Serializing a patch at index $t$ tells us exactly how to populate the patch.
In particular, each key in the patch must
be recovered to the largest $s < t$ such that $s$ is the index of a Paxos entry that updates that key.

Furthermore, patch insertion must be atomic. Otherwise, it could be interleaved with updates to keys in the patch, which would violate the safety requirement,
because then the version of the key in the patch would not reflect a prefix of $t$.
This is actually a subtle point because \ZippyDB batches many writes into a single Paxos entry, as shown in Figure~\ref{fig:patchReq}. If patch insertion is
batched with other writes, then the patch will not reflect the writes that are in front of it
in the batch.
Hence, we force the patch insertion to be its own Paxos entry.

Even though it stores a relatively small amount of data, the Paxos protocol itself can tolerate bit errors
by writing an additional entry per Paxos entry (for more information, see PAR~\cite{paxos-recovery}).

\paragraph{Local Metadata Duplication.}
There are two flavors of metadata in RocksDB: metadata files and metadata blocks in SST files.
Metadata files, such as a MANIFEST, OPTIONS, and CURRENT,
are only read during startup and then cached in memory. We can easily protect these
metadata files by locally replicating them, which
adds a minimal space overhead (on the order of kilobytes per server).
Other files such as LOG files don't need to protected, as they simply contain printed log statements used
for debugging.

Metadata blocks, however, must be protected because the integrity of the recovery
process depends on uncorrupted index blocks, and index blocks are not replicated (since each local SST file is unique).
We protect metadata blocks by writing
them several times in-line in the same SST file. In our implementation, we write each metadata block
twice\footnote{For increased protection, metadata blocks can be locally replicated more than twice
or protected with software error correction.}.
Protecting metadata enables us to isolate errors to a single data block, rather than invalidating an entire SST file.

\begin{figure}[t]
\centering
\includegraphics[trim={0.4cm 3.5cm 3cm 3.7cm},clip,scale=0.38,page=8]{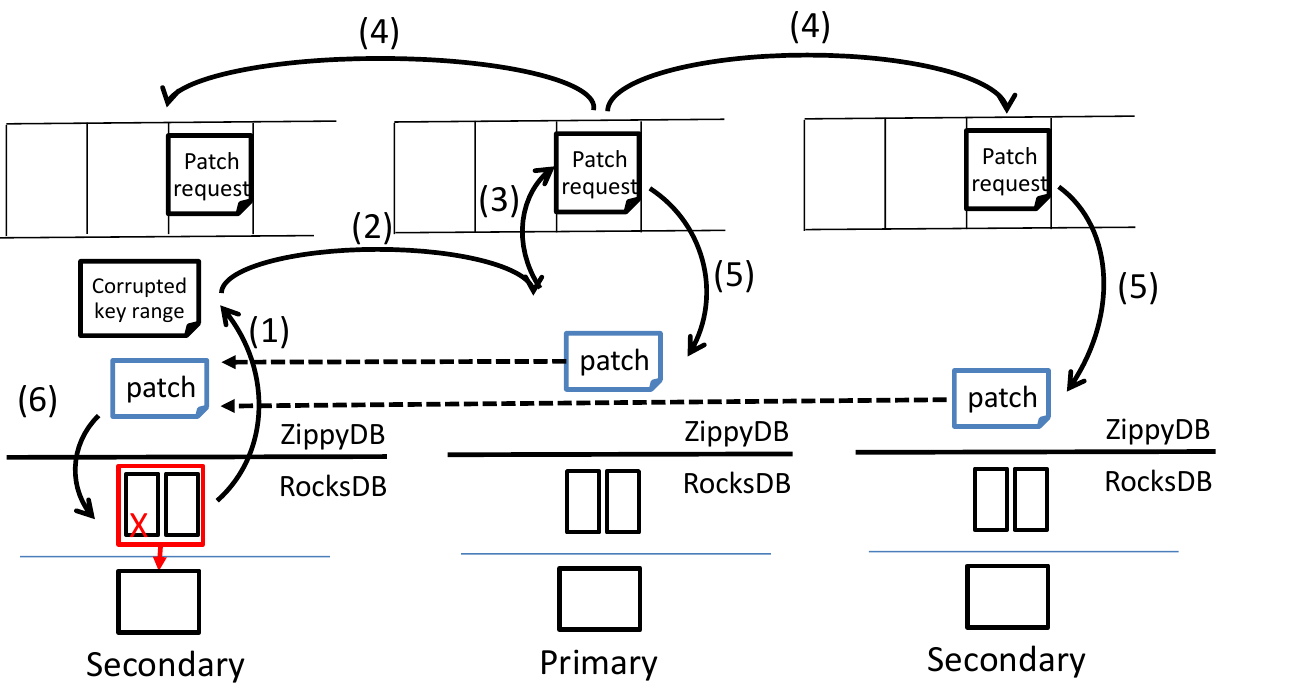}
\caption{Recovering a corrupted RocksDB data block involves the following steps:
(1) RocksDB compaction iterator determines the corrupted key range based on the index blocks of the SST files and reports this to \ZippyDB.
(2) The \ZippyDB shard reports this error to the primary for that replica.
(3) The primary shard adds the patch request to the Paxos log.
(4) The Paxos engine replicates the request to all replicas.
(5) Each replica tries to process the patch request. 
If the processing shard is \emph{not} the corrupted shard, then it prepares a patch from its local RocksDB state and sends it to
the corrupted shard.
If the processing shard \emph{is} the corrupted shard, then it waits for a patch from any of the other replicas.
(6) The corrupted shard applies the fresh patch to its local RocksDB store.}
\label{fig:zippyRecovery}
\end{figure}


\subsubsection{\system Recovery in \ZippyDB}
\label{sec:zipRecovery}
\ZippyDB does not synchronously recover corrupted blocks encountered in user reads.
Instead, it returns the error to the client, which will retry on a different replica,
and \ZippyDB will then trigger a manual compaction involving the file containing the corrupted
data block.

\ZippyDB triggers synchronous recovery only when
a corruption error occurs during compaction. Figure~\ref{fig:zippyRecovery} depicts this process.
Importantly, we do not release
a compaction's output files until the recovery procedure finishes; otherwise, stale key versions
may reappear in the key ranges still undergoing recovery.
Fortunately, because compaction is a background process, we can wait
for recovery without affecting client operations.

Step (1) is implemented entirely within RocksDB. In particular, a RocksDB compaction iterator will
record a corrupted key range when it's encountered, and then skip it to continue scanning. At the end
of the iterator's lifetime, \ZippyDB is notified about the corrupted key range. If there are
multiple corrupt key ranges, they are batched into a single patch request.

Step (3) must go through the primary because the primary is the only shard that can
propose entries to the Paxos log. Note that this does not mean primaries cannot 
recover from corrupted data blocks. The patch request that goes in the Paxos log is simply
a no-op that reserves a point of reference for the recovery procedure and includes
information necessary for recovery, such as the corrupted key ranges and the ID of
the corrupted shard. Any replica
that encounters the patch request in the log is by definition up-to-date to that point in
the Paxos log, which means any replica that isn't the corrupted replica can send a patch
to the corrupted replica.

In Step (5), an uncorrupted replica creates a patch on the affected key range
with a RocksDB iterator. Note that it might encounter a bit corruption
while assembling the patch. In practice the probability of this is very small because
the number of keys covered by the patch is on the order of kilobytes (\S\ref{hihihi}).
However, if a corruption is encountered while assembling a patch, the replica simply
does not send a patch. Therefore, for the patch request to fail,
\emph{both} (or more, if the replication factor is more than 3) uncorrupted replicas will 
have to encounter a bit corruption, and this probability is low (see Table~\ref{tab:eval_comparison}).

Step (6) is also implemented at the RocksDB level. When a replica applies a patch,
simply inserting all the key-value pairs present in the patch is insufficient because
of deleted keys. In particular, any key present in the requested key range and \emph{not} present
in the patch is an implicit delete. Therefore, to apply a patch, the corrupted shard must
also delete any keys that it can see that aren't present in the patch.
This case is possible because RocksDB
deletes keys by inserting a tombstone value, which is inlined in SST files.
Hence the corrupted data block may contain tombstone operators that
delete a key, and these must be preserved. 

\subsubsection{Invalidating Snapshots}
In RocksDB, users can request snapshots, which are represented by a sequence
number.
Then, for as long as the snapshot with sequence number $s$ is active, RocksDB will not delete
any version, $s'$, of a key where $s'$ is the greatest version of the key such that $s' < s$.
\ZippyDB uses RocksDB snapshots to execute transactions. If RocksDB invalidates
a snapshot, then the transaction using that snapshot will abort and retry.

A subtle side-effect of a corrupted data block is snapshot corruption. For example, suppose
the RocksDB store has a snapshot at sequence number 100 and the corrupted data block contains
a key with sequence number 90. 
For safety, we need to invalidate any snapshots that could have been affected by the corrupted key range.
Because the data block is corrupted, it cannot be read, so we
do not know whether this corruption affects snapshot 100.
For now, we take the obviously correct approach and invalidate all local
snapshots of the RocksDB shard affected by the corruption.
In practice, this is reasonable because most RocksDB snapshots have short lifetimes.

\subsection{HDFS-\system}
\label{sec:sleddyhdfs}
\subsubsection{HDFS Overview.}
HDFS is a distributed file system that is designed for storing large files
that are sequentially written and read.
Files are divided into 128MB blocks, and HDFS replicates and reads at the block level.

HDFS servers have three main roles: NameNode, JournalNode, and DataNode. The NameNode 
and JournalNodes 
store cluster metadata such as
the cluster directory structure and mappings from block to DataNode.
JournalNodes quorum-replicate updates to this metadata by running a protocol similar to Multi-Paxos; 
there is no leader election because the NameNode is the leader,
and HDFS deployments run a ZooKeeper service to ensure there is always one live
NameNode~\cite{hahdfs}. 

As with the Paxos log of \ZippyDB,
we can protect against bit errors in the JournalNode by adding an additional entry~\cite{paxos-recovery}.
To prevent the JournalNode logs from growing indefinitely,
the NameNode takes periodic snapshots 
of the stored metadata.
We divide the snapshots into 512 byte chunks
and compute a CRC32 checksum for each chunk,
just as with data blocks. 
During NameNode recovery, which runs only during recovery or startup mode
and not during the steady-state, snapshot corruptions can be
fixed by fetching the corresponding chunk from the standby NameNode,
which acts as a hot NameNode backup.

DataNodes store actual HDFS data blocks (they are the local data stores
in Figure~\ref{fig:everything}),
and they respond to client requests to read blocks.
If a client encounters errors while reading a block, 
it will continue trying other DataNodes from the offset of the error
until it can read the entire block.
Once it encounters an error on a DataNode, the client will not try that node again.
If there are no more DataNodes and the block is not fully read, the read fails
and that block is considered missing.

Additionally, HDFS has a configurable background ``block scanner'' that periodically scans
 data blocks and reports corrupted blocks for re-replication.
But the default scan interval
is three weeks, and even if the periodic scan does catch bit errors
before the next read of a block, the NameNode can only recover at the 128 MB 
block granularity.
If there is a bit error in every replica of a block, then HDFS cannot recover
the block.

\subsubsection{Implementing \system}
\paragraph{Reducing Error Amplification of Data Blocks}
\label{sec:datanodes}
We leverage the
observation that
HDFS checksums every 512 bytes in 
each 128 MB data block.
Corruptions thus can be narrowed down to a 512 byte chunk;
verifying checksums adds no overhead,
because by default HDFS will verify checksums during every block read.
For streaming performance, 
the smallest-size buffer that is streamed during a data block read is 64 KB,
so we actually repair 64 KB everytime there is a corruption.
To mask corruption errors from clients, we repair a data block
synchronously during a read.
Under \system, the full read (and recovery) protocol is the following.

\begin{figure}[t]
\begin{subfigure}[t]{\columnwidth}
\centering
\includegraphics[trim={5cm 9.4cm 10cm 6.5cm},clip,scale=0.7,page=11]{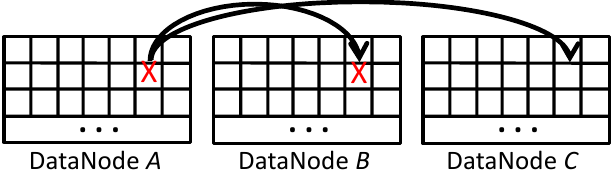}
\caption{$A$ will attempt to correct a corrupted chunk
by requesting it from other datanodes until it receives a clean chunk.}
\label{fig:hdfsrec}
\vspace{0.1in}
\end{subfigure}
\hspace{0.1in}
\begin{subfigure}[t]{\columnwidth}
\centering
\includegraphics[trim={5cm 2.7cm 10cm 12cm},clip,scale=0.5,page=11]{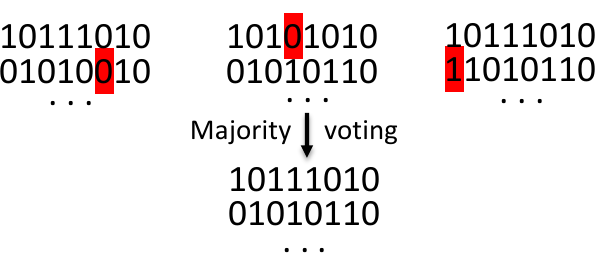}
\caption{If the chunk is corrupted on all datanodes, then $A$ will attempt
majority voting to reconstruct a clean chunk.}
\label{fig:majVote}
\end{subfigure}
\caption{DataNodes will stream the read of a block in 64 KB chunks. When it encounters
a checksum failure, a DataNode will try to repair the individual chunk.}
\end{figure}

Each 128 MB block in HDFS is replicated on three
DataNodes, call them $A,B,C$.
An HDFS read of a 128 MB block is routed to one of these
DataNodes, say $A$. 
$A$ will stream the block to the client in 64 KB chunks,
verifying checksums before it sends a chunk.
If there is a checksum error in a 64 KB chunk, then $A$
will attempt to repair the chunk by requesting the 64 KB chunk
from $B$. If the chunk sent by $B$
also contains a corruption, then the checksum will be incorrect,
and $A$ will request the chunk from $C$ (see Figure~\ref{fig:hdfsrec}).

If $C$ \emph{also}
sends a corrupted chunk, then $A$ will attempt to construct a
correct version of the chunk through
bit-by-bit majority voting: the value of a bit in the chunk is the majority vote 
across the three versions provided by $A$, $B$, and $C$.
The idea behind majority voting is that the probability that the
corruptions on $A$, $B$, and $C$ affect the same
byte is very low, which means a majority vote across the three versions
of the byte should end up with the correct data.
After reconstructing the chunk via majority voting (Figure~\ref{fig:majVote}), $A$ will verify 
the checksums again; if the checksums fail, then the read fails.
As we show in Section~\ref{sec:evalHDFS}, UBERs
have to be at least $10^{-8}$ in order for majority voting failures
to affect read failures, which allows HDFS-\system to tolerate on the order of a
million times more bit errors than HDFS.

Note that bit-by-bit majority voting is possible only if
the device can return pages with uncorrectable errors (see \S\ref{sec:device-support});
otherwise, our HDFS implementation simply uses chunk-by-chunk
recovery.
Furthermore, for majority voting to add significant recovery power over
chunk-by-chunk recovery, the number of corrupt
bits returned by the device should be relatively small compared to the page size;
the number of corrupt bits on a device page after running hardware ECC is dependent on the ECC function and its implementation.

\paragraph{Safe Recovery Semantics.}
Safety is straightforward in HDFS
because data blocks are immutable once written, so there are never in-place updates
that will conflict with chunk recovery.
Before a client does a block read, it first contacts the NameNode to get
the DataNode IDs of all the DataNodes on which the block is replicated.
When a client sends a block read request to a DataNode, it also
sends this set of IDs.
Because blocks are immutable, these IDs are guaranteed to be correct replicas of
the block, \emph{if they exist}. It could be that a concurrent operation has deleted the block.
In this case, if chunk recovery cannot find the block 
on another DataNode because it has been deleted, then it cannot perform recovery,
so it will return the original checksum error to the client. This is correct, because there is no guarantee in HDFS
that concurrent read operations should see the instantaneous deletion of a block.

\paragraph{Local Metadata Duplication.}
Each role in HDFS has local metadata files that must be correct, otherwise the role cannot be
started. These files include a VERSION file, as well as special files on the NameNode and JournalNode.
For example, the NameNode
stores a special file (\texttt{seen-txid}) which contains a high-water
mark transaction ID. Any correct recovery of the existing cluster must be able to recover 
up to at least this transaction.

Metadata files are not currently protected in HDFS; thus, a single
corruption will prevent the role from starting. 
To implement \system, we add a standard CRC32 checksum
at the beginning of each file and replicate the file twice so that there are actually three copies of the file
on disk.
If there is a checksum error when the file is read, the recovery protocol will visit each of the copies until
it finds one with a correct checksum.

\section{Evaluation}
This section addresses the three following questions.  (1)
What is the highest UBER that \ZippyDB and HDFS can tolerate with \system?
(2) How is \ZippyDB's recovery time affected by \system?
(3) What are the overheads of \system on steady-state requests in HDFS?

\paragraph{Experimental Setup.}
To evaluate \ZippyDB, we set up a cluster of \company servers that capture
and duplicate live traffic from a heavily loaded service used in computing
user feeds.
To evaluate
HDFS, we run experiments on a cluster of 10 machines (each
with a role described below) each with
8 ARMv8 cores at 2.4~GHz, 96~GB of RAM, and 120~GB of flash.
In the cluster, we allocate
one machine each for a NameNode, standby NameNode, and JournalNode,
and three machines run the DataNode role. Four machines act as HDFS clients.
HDFS experiments have a load and read phase: in the load phase, we
load the cluster with 200, 128MB files with random data. In the read phase, clients
randomly select files to read. After the load phase, we clear the page cache.

\begin{figure*}
\centering
\begin{subfigure}{0.475\textwidth}
\includegraphics[width=\textwidth]{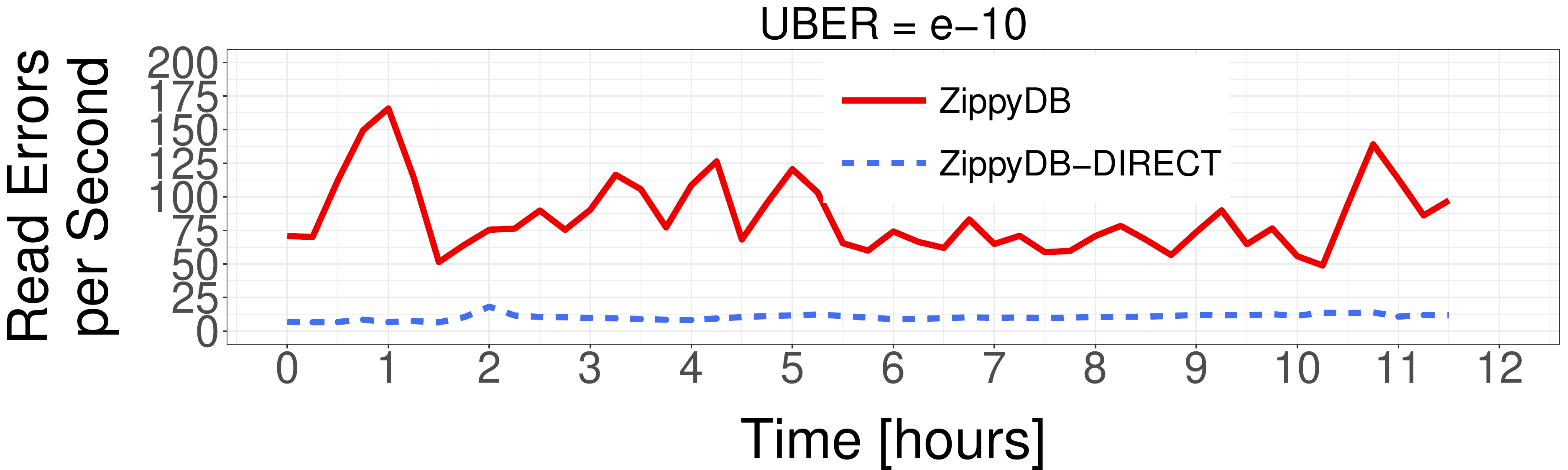}
\label{fig:readErrors10}
\end{subfigure}
\begin{subfigure}{0.475\textwidth}
\includegraphics[width=\textwidth]{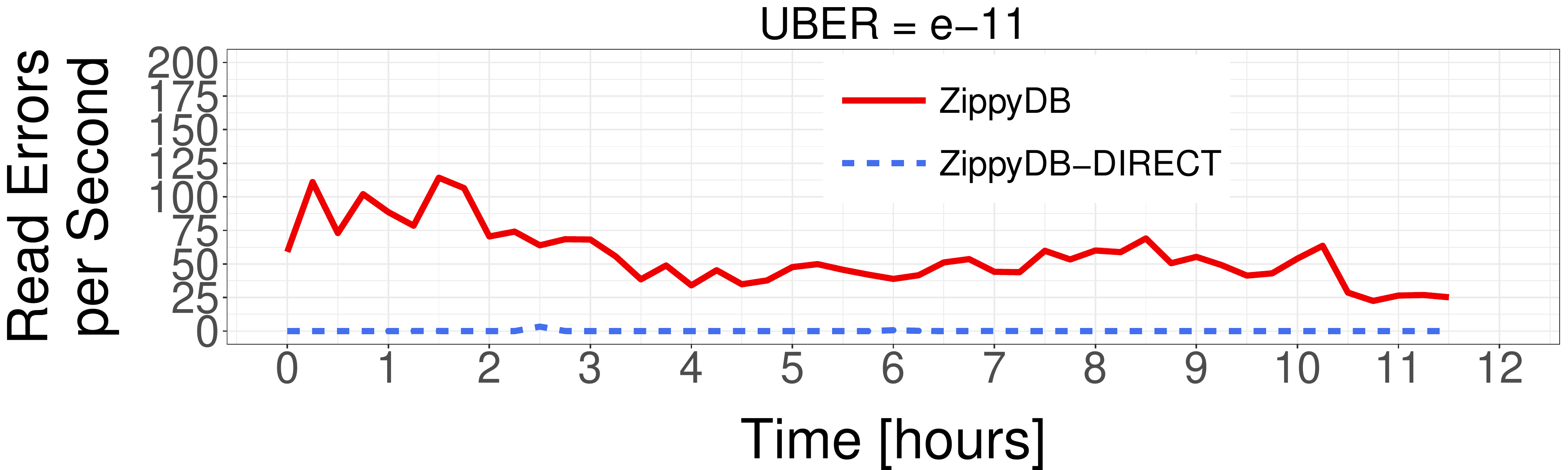}
\label{fig:readErrors11}
\end{subfigure}
\begin{subfigure}{0.475\textwidth}
\includegraphics[width=\textwidth]{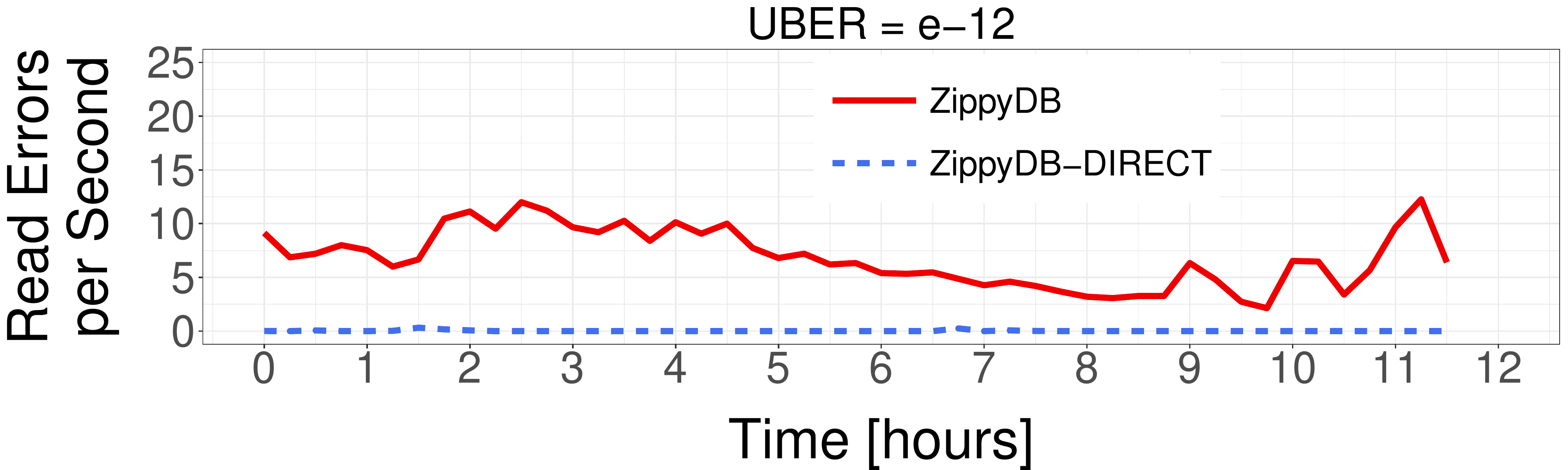}
\label{fig:readErrors12}
\end{subfigure}
\begin{subfigure}{0.475\textwidth}
\includegraphics[width=\textwidth]{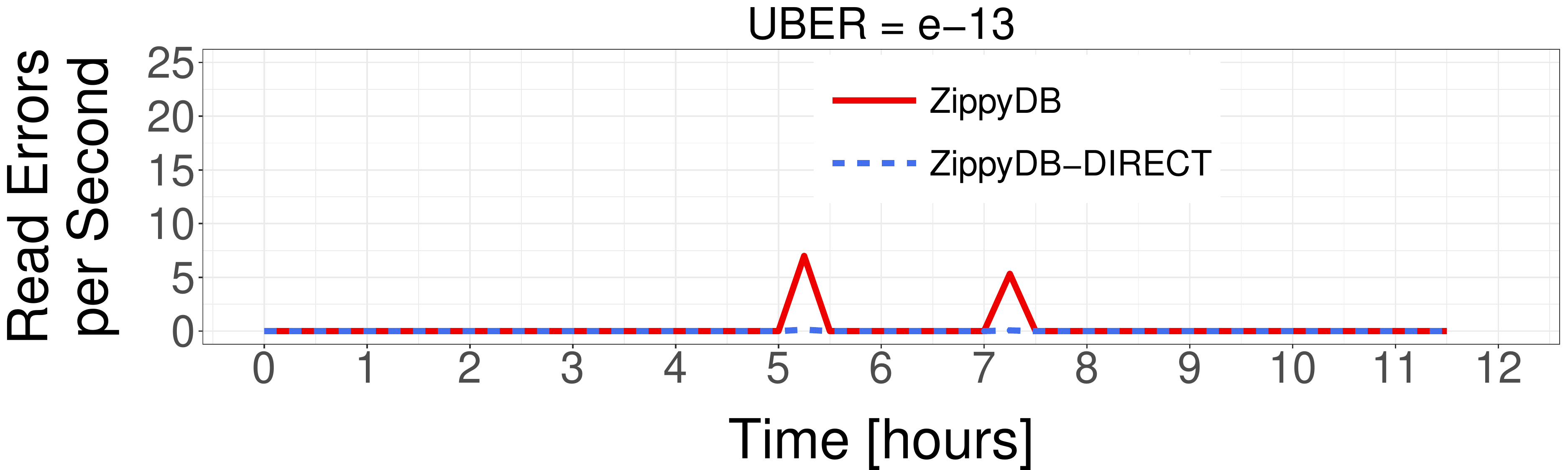}
\label{fig:readErrors13}
\end{subfigure}
\caption{Read error rates over time in \ZippyDB and \ZippyDB-\system, for a variety of UBERs.}
\label{fig:readErrors}
\end{figure*}

\paragraph{Error Injection.}
To simulate UBERs, we inject bit errors into the files of both systems.
In \ZippyDB, we inject errors with a custom RocksDB environment that
flips bits as they are read from a file.
In HDFS, we run a script in between the load and read phases
that flips bits in on-disk files and flushes them.
For an UBER of $\mu$, e.g. $\mu = 10^{-11}$, we inject errors at the rate of 1 bit flip
per $1/\mu$ bits read.
We tested with UBERs higher than the manufacturer advertised $10^{-15}$
to test the system's performance under high error rates, and so that we can
measure enough bit errors during an experiment time of 12 hours rather than several days (or years)\footnote{
Note that an UBER $10^{-11}$ is 10,000$\times$ \emph{higher} than $10^{-15}$}.

\subsection{\ZippyDB}
\paragraph{UBER Tolerance.}
One main difference between unmodified \ZippyDB and \ZippyDB-\system is
that \ZippyDB-\system avoids crashing when encountering a bit error. 
To characterize how many server crashes are mitigated with \system,
we measured the average rate of compaction errors per hour \emph{per server}, over 12 hours.
The results are shown in Table~\ref{tab:num_compact_errs}.
Figure~\ref{fig:readErrors} shows the read error rate over time of both
systems for a variety of UBERs.
Note that the error rate patterns across UBERs are different because
they are run during different time intervals, so each UBER experiment
sees different traffic.
The error rate is much higher for \ZippyDB than \ZippyDB-\system because not only do clients 
see errors from regular read operations, but also they experience
the spike in errors when a server shuts down due to a compaction corruption.
This is true across the range of evaluated UBERs.

\begin{table}[t!]
  \centering
  \small
  \begin{tabular}{cc}
    \toprule
    UBER & Compaction Errors per Hour per Server \\
    \midrule
    $10^{-10}$ & $0.1991\pm 0.1077$  \\
    $10^{-11}$ & $0.0621\pm 0.0455$  \\
    $10^{-12}$ & $0.0038\pm 0.0035$  \\
    $10^{-13}$ & $0.0003\pm 0.0005$  \\
    \bottomrule
  \end{tabular}
  \caption{Number of compaction errors encountered by \ZippyDB. \ZippyDB-\system is able
to fix these errors, while the server crashes in \ZippyDB.}
  \label{tab:num_compact_errs}
\end{table}

\begin{figure}[t]
\centering
\includegraphics[scale=0.22]{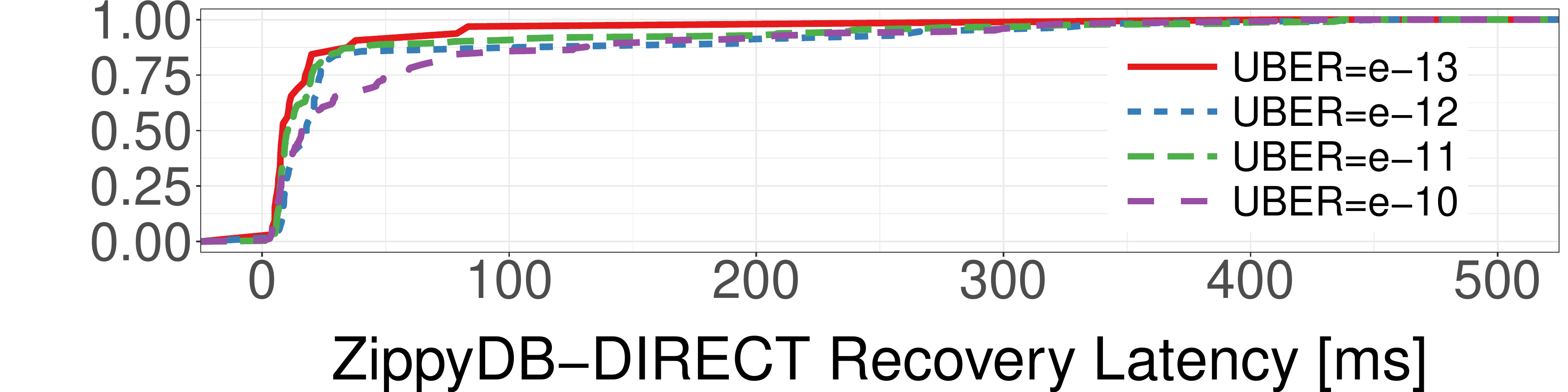}
\caption{CDF of compaction recovery latencies in \ZippyDB-\system. \ZippyDB-\system takes milliseconds
to recover from corruptions, while \ZippyDB takes \emph{minutes}.}
\label{fig:compactRecLat}
\end{figure}

\begin{figure}[t]
\centering
\includegraphics[scale=0.22]{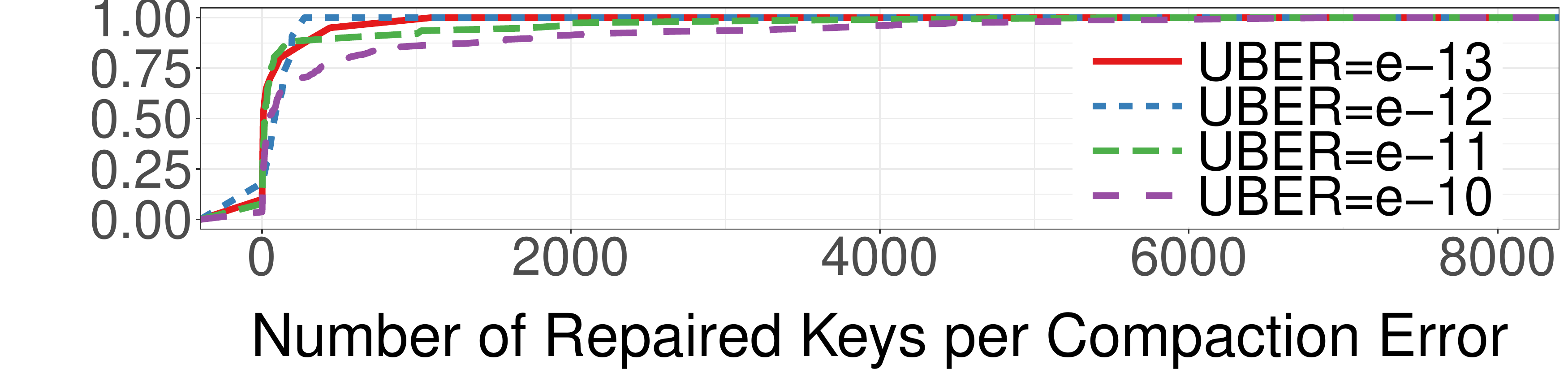}
\caption{CDF of patch sizes generated during the \ZippyDB-\system recovery process.
The patch size is small, which means low error amplification.}
\label{fig:diffSize}
\end{figure}

\paragraph{Time Spent in Reduced Durability.}
\label{hihihi}
With \system, we also seek
to minimize the amount of time spent in reduced durability
to decrease the likelihood of simultaneous replica failures.
Figure~\ref{fig:compactRecLat} shows a CDF of the time it takes 
to recover from compaction errors
in \ZippyDB-\system. The graph shows the amount of time it takes
for replicas to process the Paxos log up until the patch request, as well as the overhead
of constructing and inserting the patch.
With \system, this recovery time is on the order of \emph{milliseconds}.

In contrast,
the period of reduced durability in unmodified \ZippyDB due to a compaction error is
on the order of \emph{minutes}, depending on the amount of data stored in the
crashed \ZippyDB server.
This is directly due to the high error amplification of \ZippyDB,
which invalidates 100s of RocksDB shards due to a single compaction bit error.
With \system, \ZippyDB can reduce its recovery time due to a bit error by around 
10,000x!

We also found that the recovery latency is dependent on the size
of the patch required to correct the corrupted key range.
Figure~\ref{fig:diffSize} presents a CDF of the size of the patches generated during the
recovery process. 
Patch size is also interesting because the recovery mechanism described in Section~\ref{sec:zipRecovery}
recovers a \emph{range} of keys, since the exact keys on the corrupted 
data block are impossible to identify. As we see in Figure~\ref{fig:diffSize},
even though recovering a range can in theory increase error amplification,
the number of keys required for recovery is still low.

Figure~\ref{fig:diffSize} also confirms that as the UBER increases,
patch sizes increase due to more key ranges getting corrupted during
a single compaction operation.

\paragraph{Reduced CPU Consumption.}

\begin{table}[t!]
  \centering
  \small
  \begin{tabular}{p{0.6cm}p{3cm}p{3cm}}
    \toprule
    UBER & \ZippyDB (CPU consumption) & \ZippyDB-\system (CPU consumption) \\
    \midrule
    $10^{-10}$ & 100\% & 80\%  \\
    $10^{-11}$ & 100\% & 51\% \\
    $10^{-12}$ & 100\% & 51\%  \\
    \bottomrule
  \end{tabular}
  \caption{CPU consumption of \ZippyDB and \ZippyDB-\system (lower is better),
   normalized to \ZippyDB.}
  \label{tab:cpu}
\end{table}

Due to its more efficient recovery from bit corruptions, \ZippyDB-\system consumes
much less CPU than \ZippyDB, as shown in Table~\ref{tab:cpu}.
We don't report statistics for UBER = $10^{-13}$ because the errors
are infrequent. CPU usage is higher in \ZippyDB mostly due to handling redirected
client requests as well as shard restarts.

\subsection{HDFS}
\label{sec:evalHDFS}
\begin{figure}[t]
\centering
\includegraphics[width=\columnwidth]{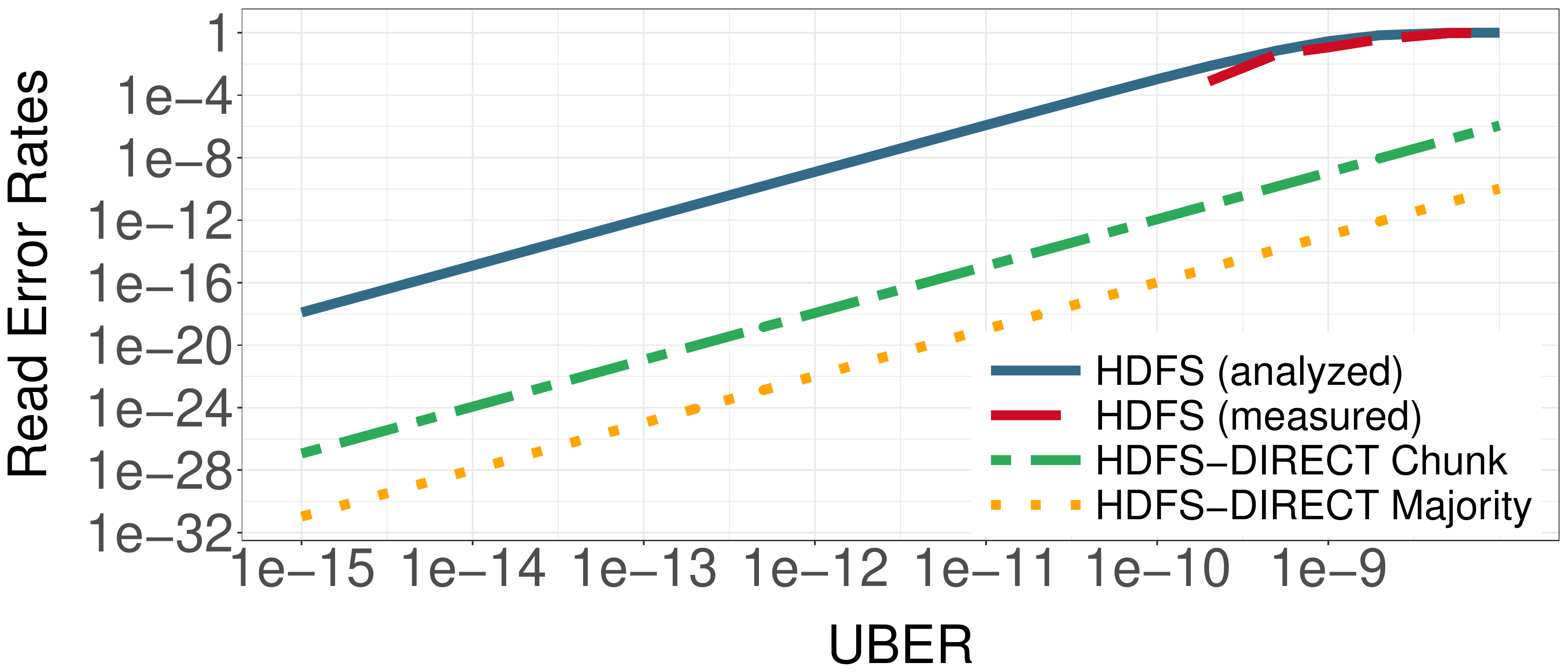}
\caption{Read error rate for HDFS with varying UBER.
The HDFS (analyzed), HDFS-\system Chunk and HDFS-\system Majority are all computed
using the formula in \S\ref{sec:availability-analysis}. HDFS-\system Chunk
is based on chunk-by-chunk recovery, while HDFS-\system Majority is computed
on bit-by-bit majority. Bit-by-bit majority provides lower error rates
due to its lower recovery amplification. HDFS (Measured) is
the measured HDFS read errors. 
With HDFS-\system we could not measure any level of read errors until
UBERs of $10^{-4}$.}
\label{fig:HDFSerrors}
\end{figure}

\paragraph{UBER Tolerance.}
The main advantage of HDFS-\system over HDFS is the ability to tolerate much
higher UBERs with chunk-level recovery.
Figure~\ref{fig:HDFSerrors} reports block read error rates
of HDFS with varying UBERs.
This read error is also considered \emph{data loss} in
HDFS, because the data is unreadable (and hence unrecoverable) even after
trying all 3 replicas.
The figure shows both the measured read error on our HDFS experimental setup,
as well as the computed read error based on the computation presented in \S\ref{sec:availability-analysis}.
The experimental read error is collected by running thousands of file reads
and measuring how many fail.
Within the UBER range in which we could effectively measure errors,
the read errors we measured were similar to the computed results.
We do not present experimental read error rates for HDFS-\system, because the read
error rates are too low to be measured for the UBERs tested in Figure~\ref{fig:HDFSerrors}.
The figure also presents the expected error rates for HDFS-\system using
chunk-by-chunk recovery and bit-by-bit majority. As expected, bit-by-bit majority reduces
the read error rate due to its lower error amplification (it can recover bit-by-bit).
Both our analysis and the experimental results show that
HDFS-\system can tolerate a 10,000-100,000x higher UBER and maintain the same read
error rate!

\paragraph{Overhead of \system.}
Table~\ref{tab:HDFSthruput} shows the throughput of both systems,
measured by saturating the DataNodes with
four, 64-threaded clients that are continuously reading random files.
The throughput of HDFS goes to zero at an UBER of $10^{-8}$,
because it cannot complete any reads due to corruption errors.
Such failures do not occur in 
HDFS-\system, although its throughput decreases modestly 
as UBER increases due to the
overhead of synchronously repairing corrupt chunks during reads.

\begin{figure}[t]
\centering
\hspace{0.05in}
  \begin{subfigure}{0.5\textwidth}
    \includegraphics[width=0.95\textwidth]{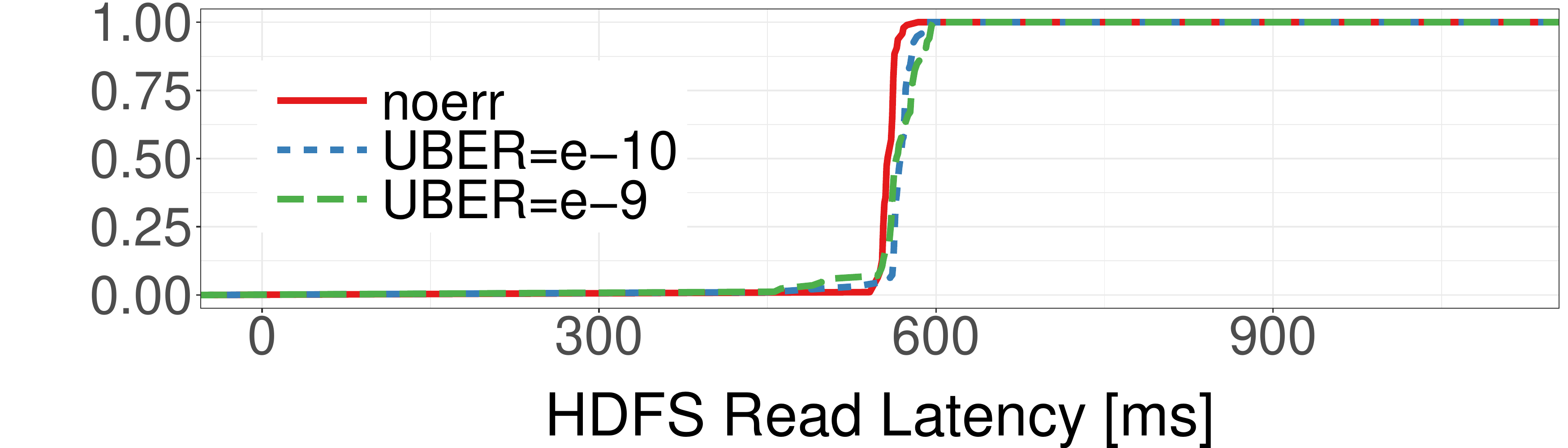}
    \label{fig:hdfs_latency_hdfs}
  \end{subfigure}
  \begin{subfigure}{0.5\textwidth}
    \includegraphics[width=0.95\textwidth]{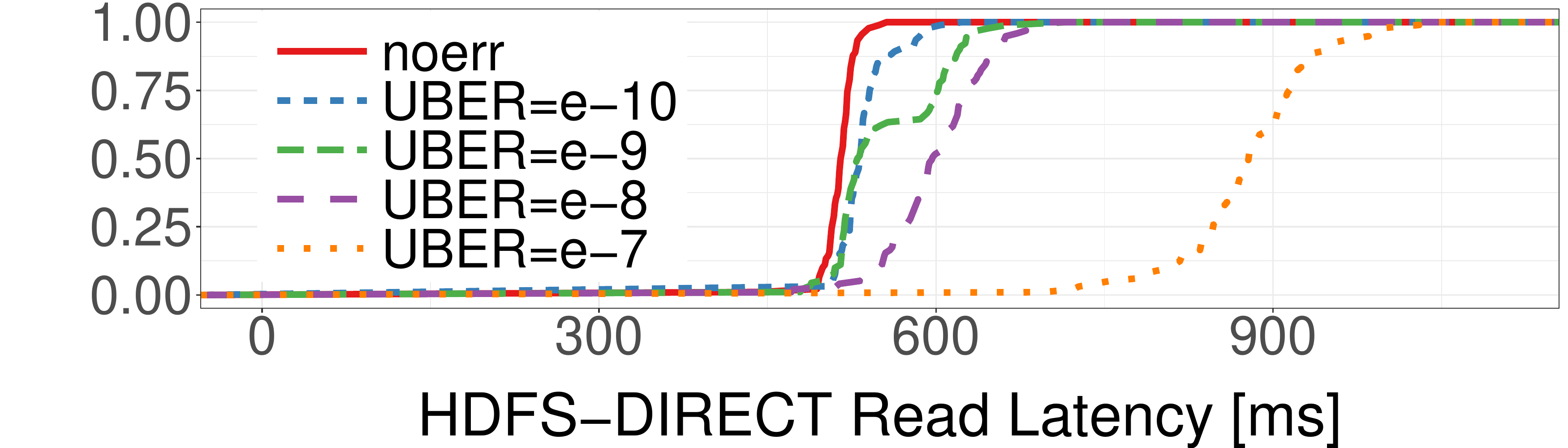}
  \end{subfigure}
  \caption{Read latencies (128~MB) of HDFS and HDFS-\system. 
All reads fail in HDFS an UBER of $10^{-8}$ and higher.}
  \label{fig:hdfs_latencies}
\end{figure}

For HDFS-\system, we are also interested in latency incurred by synchronous
chunk recovery.
We compare the CDF of read latencies of 128~MB blocks for different UBERs
in Figure~\ref{fig:hdfs_latencies}.
The higher the UBER, the more chunk recovery requests that need to be made during
a block read and the longer these requests will take. The results in
Figure~\ref{fig:hdfs_latencies} (and Table~\ref{tab:HDFSthruput})
highlight the fine-grained tradeoff between performance
and recoverability that is exposed by \system.
We also report HDFS read latencies, but there is little difference across UBERs
because only latency for successful block reads are included.
Note that the CDF for HDFS does not 
include UBERs higher than $10^{-8}$, since at those error rates HDFS cannot 
read a block without an error.

\begin{table}[t!]
  \centering
  \small
  \begin{tabular}{p{1.3cm}p{2.5cm}p{2.7cm}}
    \toprule
    UBER & HDFS throughput [GB/s] & HDFS-\system throughput [GB/s] \\
    \midrule
    $10^{-7}$ & $0.00\pm0.00$ & $2.09\pm0.08$  \\
    $10^{-8}$ & $0.00\pm0.00$ & $2.56\pm0.09$  \\
    $10^{-9}$ & $2.46\pm0.08$ & $2.55\pm0.07$  \\
    $10^{-10}$ & $2.89\pm0.10$ & $2.84\pm0.07$ \\
    No errors & $2.83\pm0.07$ & $2.88\pm0.07$ \\
    \bottomrule
  \end{tabular}
  \caption{Throughput of HDFS and HDFS-\system. At UBER$=10^{-8}$,
  HDFS throughput collapses due to bit errors.}
  \label{tab:HDFSthruput}
\end{table}

\section{Discussion}
\label{sec:discussion}
\paragraph{Local File System Error Tolerance.}
\label{sec:file-systems}

Distributed storage systems run on top of local file systems.
Therefore, when devices exhibit higher UBERs, local file systems also experience
higher UBERs.
\system protects application-level metadata and data, which are
just data blocks at the local file system level.
Protecting local file system metadata (such as inodes, the FS journal, etc.)
is beyond the scope of this paper.
Several existing file systems protect metadata against bit corruptions~\cite{IRON,ZFS,xfschecksum,refs,NOVA,FlexFS,Gunawi:2007}.
The general approach is to add checksums to file system metadata and
locally replicate it for error correction.
Another approach
is to use more reliable hardware for metadata, and less reliable hardware
for data blocks~\cite{FlexFS}.

\paragraph{Support for \system.}
\label{sec:device-support}
\system does not require any hardware support.
However, a couple of simple device-level mechanisms would help datacenter operators
run devices past their manufacturer defined UBER.
First, it would be beneficial if devices have a less aggressive ``bad block policy'', which is a
firmware protocol for retiring blocks 
once they reach some heuristic-defined level of errors.
Second, it would be beneficial if devices
return the content of pages, even if they have an error. This enables
distributed storage applications to minimize their recovery amplification, since they
can recover data at a granularity smaller than a device page (e.g., on a bit-by-bit
level using majority voting). This is not a hard requirement, since as we showed in \S\ref{sec:availability-analysis}
even recovering at a device page level (e.g., 4-8~KB) provides significant benefits.
In case corrupt pages cannot be read, it is important
to guarantee that when duplicating metadata the copies are stored on separate physical pages.
Otherwise, a page error could invalidate all copies of the metadata.

\section{Related Work}

Related work is divided into two main parts: systems that deal with device
errors using software mechanisms or by applying more aggressive hardware mechanisms.

\paragraph{Software-level Redundancy.}
\system is related to Protocol Aware Recovery (PAR)~\cite{paxos-recovery}, which recently
demonstrated how consensus-based protocols can be adapted to address bit-level errors.
Unlike PAR, which only addresses consensus protocols, our work tackles
bit-level errors in general purpose storage systems. We also show how increasing
the resiliency to bit-level errors can significantly reduce storage costs and improve
live recovery speed in datacenter environments.

FlexECC~\cite{flexECC} and Duracache~\cite{duracache} are flash-based key-value caches
that use less reliable disks by treating devices errors as cache misses.
D-GRAID is a RAID storage system that gracefully degrades
by minimizing the amount of data needed to recover
from bit corruptions~\cite{D-GRAID}.
There is a large number of distributed storage systems that use inexpensive, unreliable hardware, while providing
consistency and reliability guarantees~\cite{FAWN,GFS,dynamo}.
However, these systems treat bit corruptions similar
to entire-node failures and suffer from high recovery amplification. 

There is a large body of work on finding errors in the way both local file systems and 
distributed file systems handle disk corruptions~\cite{redundancy}. These efforts are orthogonal to our work,
because they focus on correctness flaws of existing systems under disk corruptions, while
we focus on how far we can push disk error rates without compromising performance (while maintaining correctness).
Research on hardening local file systems to tolerate disk errors supports our vision of 
less reliable disks, because it shows that it is possible to protect a local file system from disk
bit errors~\cite{IRON,ZFS,xfschecksum,refs,NOVA,FlexFS,Gunawi:2007}.

\paragraph{Hardware-level Redundancy.}
Several studies explore extending SSD lifetime via more aggressive
or adaptive hardware error correction.  Tanakamuru \emph{et al.}~\cite{Tanakamaru20112} propose
adapting codeword size based on the SSD's dynamic device wear level to
improve SSD lifetime.  Cai \emph{et al.}~\cite{cai2015data} and Liu
\emph{et al.}~\cite{retentionrelaxation} introduce techniques to
dynamically learn and adjust the cell voltage levels based on
retention age. Zhao \emph{et al.}~\cite{LDPC-SSD} propose
using the soft information with LDPC error correction to increase lifetime. 
Our approach is different: instead of improving hardware-based error correction,
we leverage existing software-based redundancy to address bit-level errors.

\section{Conclusion}
This paper presents \system, a set of policies that use
the inherent redundancy that exists in distributed storage
applications for live recovery of bit corruptions.

We can extend the approach of handling error correction in the
distributed storage layer in several directions.
First, distributed storage systems can control the level of
error correction depending on data type. For example,
some data types may be more sensitive to bit corruptions (e.g.,
critical metadata), while others may not. 
Second, distributed storage system can
control hardware mechanisms that influence the performance of the device.
For example, storing fewer bits per cell
generally reduces the latency of the device (at the expense of its capacity).
Certain applications may prefer for to use a hybrid of low latency and low capacity devices
for hot data, while reserving the high capacity devices for colder data.

\footnotesize \bibliographystyle{abbrv}
\bibliography{bib}


\end{document}